\documentclass[useAMS,usenatbib]{mn2e}

\usepackage{graphicx,amssymb,amsmath}

\title[The mass function of 35 Galactic GCs]{The Global Mass Functions of 35
Galactic globular clusters: II. Clues on the Initial Mass Function and Black Hole Retention Fraction}
\author[Baumgardt \& Sollima]{H. Baumgardt$^{1}$\thanks{E-mail:
h.baumgardt@uq.edu.au}, A. Sollima$^{2}$\\
$^{1}$ School of Mathematics and Physics, University of Queensland, St Lucia,
QLD 4072, Australia\\
$^{2}$ INAF Osservatorio Astronomico di Bologna, via Gobetti 93/3, Bologna,
40129, Italy\\
}
\begin{document}


\pagerange{\pageref{firstpage}--\pageref{lastpage}} \pubyear{2017}

\maketitle

\label{firstpage}

\begin{abstract}
In this paper we compare the mass function slopes of Galactic globular clusters recently determined by Sollima \& Baumgardt (2017) with a  
set of dedicated $N$-body simulations of star clusters containing between $65,000$ to $200,000$ stars. We study
clusters starting with a range of initial mass functions (IMFs), black hole retention fractions and orbital parameters
in the parent galaxy. We find that the present-day mass functions of globular clusters
agree well with those expected for star clusters starting with Kroupa or Chabrier IMFs, and are incompatible 
with clusters starting with single power-law mass functions for the low-mass stars. The amount of mass segregation seen in 
the globular clusters studied by Sollima \& Baumgardt (2017) can be fully explained by two-body relaxation driven 
mass segregation from initially unsegregated star clusters. Based on the
present-day global mass functions, we expect that a typical globular cluster in our sample has lost about 75\% of its mass since
formation, while the most evolved clusters have already lost more than 90\% of their initial mass and should dissolve 
within the next 1 to 2 Gyr. Most clusters studied by Sollima \& Baumgardt also show a large difference between their 
central and global MF slopes, implying that the majority of Galactic globular clusters is either near or already past core collapse. 
The strong mass segregation seen in most clusters also implies that only a small fraction of all black holes formed in 
globular clusters still reside in them.
\end{abstract}


\section{Introduction}
\label{intro_sec}

This is the second of two papers in which we explore the present-day stellar mass functions
of Galactic globular clusters. In the first paper \citep{sollimabaumgardt2017} we derived completeness corrected
stellar mass functions within the central 1.6' of 35 Galactic globular clusters based on {\it HST/ACS} data 
obtained as part of the Globular Cluster ACS Treasury 
Project \citep{sarajedinietal2007}. 
We also derived the global mass functions, structural parameters and dark remnant fractions of the studied clusters
by modeling their observed mass functions, velocity dispersion profiles and surface density profiles with isotropic,
multi-mass King-Michie models \citep{gg1979, sollimaetal2012}. Our results showed that the derived global mass functions could generally
be well described by single power-law mass functions in the mass range $0.2 < m/M_{\odot} < 0.8$ except for
the least evolved clusters. We also found a tight anti-correlation between the present-day mass functions slope
and the half-mass relaxation time of the clusters. In addition, we found that the mass fraction of dark remnants in a cluster
correlates with the mass function slope of the cluster, in the sense that clusters with flatter mass
functions have a higher remnant fraction.

In the present paper we investigate what the results obtained in \citet{sollimabaumgardt2017} imply for the stellar 
mass function with which globular clusters were born and for their subsequent evolution. To this end we compare
the observational data with a set of 16 $N$-body simulations of star clusters starting with different initial
mass functions, particle numbers, half-mass radii, orbits in their parent galaxy, and black hole (BH) retention fractions.
In addition, we also use data from the large grid of 900 $N$-body simulations recently published by \citet{baumgardt2017}. 
Our paper is organised as follows: In sec.~2 we describe our $N$-body simulations in greater detail.
In sec.~3 we compare the mass functions of the simulated clusters with the observed mass functions of Galactic 
globular clusters and in sec.~4 we draw our conclusions.

\section{Description of the $N$-body runs}
\label{nbody_sec}

The simulations in this paper were made using the GPU-enabled version of the collisional $N$-body code NBODY6 \citep{aarseth1999,nitadoriaarseth2012}.
Clusters started with particle numbers between $N=65,536$ to $N=200,000$ stars. 
The initial mass functions (IMFs) of the clusters were given by either \citet{kroupa2001}, \citet{chabrier2003} or \citet{salpeter1955}
mass functions. In all simulations, stars were distributed between mass limits $0.1 < m <100$ M$_\odot$ and stellar evolution
was modeled by the stellar evolution routines of \citet{hurleyetal2000}, assuming a metallicity of $[Fe/H] = -1.30$. This metallicity
is close to the average metallicity of Galactic globular clusters. We assumed a retention fraction of neutron stars of 10\% in our simulations. 
Neutron stars not retained in the simulation were given large enough kick velocities upon formation so that they leave the clusters. In most 
simulations we also assumed a retention fraction of 10\% for the black holes,
however we also made simulations of star clusters with either 30\%, 50\% or 100\% black hole retention fractions to test the influence
of the black hole retention fraction on the cluster evolution. The initial half-mass radii of the clusters were equal to either $r_h=1$ pc, 2 pc or 4 pc
to simulate the evolution of star clusters starting with different initial relaxation times. 
All clusters followed \citet{king1966} density profiles initially with a dimensionless
central potential $W_0=5.0$.
\begin{table*}
\caption[]{Details of the performed $N$-body runs.}
\begin{tabular}{rrlcrcrrccrcr}
\noalign{\smallskip}
\multicolumn{1}{c}{Model} & \multicolumn{1}{c}{$N$}& \multicolumn{1}{c}{Mass} & \multicolumn{1}{c}{$R_G$} &
\multicolumn{1}{c}{$M_0$} & \multicolumn{1}{c}{$r_h$} & \multicolumn{1}{c}{$f_{Bin}$} & \multicolumn{1}{c}{BH ret.} &
 \multicolumn{1}{c}{$r_t$} & \multicolumn{1}{c}{$t_{rh}$} & \multicolumn{1}{c}{$T_{Diss}$} & $\alpha_g$ & $\alpha_i$ \\
\multicolumn{1}{c}{Nr.} & & \multicolumn{1}{c}{Function} & \multicolumn{1}{c}{[pc]} & \multicolumn{1}{c}{[$M_\odot$]} & [pc] & & & [pc] & \multicolumn{1}{c}{[MYR]} & \multicolumn{1}{c}{[MYR]} & & \\
\noalign{\smallskip}
1 &  65536 & Kroupa   &  8500 &  40815 & 2.0 & 0 & 10 & 50.79 & 219.0 & 20520 & $-0.52 \pm 0.01$  & $0.13 \pm 0.02$ \\
2 &  65536 & Chabrier &  8500 &  42710 & 2.0 & 0 & 10 & 51.56 & 206.8 & 20110 & $-0.50 \pm 0.02$  & $0.26 \pm 0.03$ \\
3 &  65536 & Salpeter &  8500 &  25021 & 2.0 & 0 & 10 & 43.14 & 353.1 & 16450 & $-0.63 \pm 0.02$  &$ 0.37 \pm 0.04$ \\[+0.1cm]
4 & 131072 & Kroupa   &  5000 &  83853 & 2.0 & 0 & 10 & 45.33 & 282.2 & 20750 & $-0.56 \pm 0.01$  & $0.09 \pm 0.03$ \\
5 & 131072 & Chabrier &  5000 &  89355 & 2.0 & 0 & 10 & 46.30 & 270.1 & 20540 & $-0.91 \pm 0.01$  & $0.15 \pm 0.03$ \\
6 & 131072 & Salpeter &  5000 &  49569 & 2.0 & 0 & 10 & 38.04 & 364.5 & 16510 & $-0.85 \pm 0.01$  & $0.30 \pm 0.04$ \\[+0.1cm]
7 & 131072 & Kroupa   &  3500 &  83853 & 2.0 & 0 & 10 & 32.24 & 282.2 & 13420 & $ 0.51 \pm 0.02$  & $1.35 \pm 0.04$ \\
8 & 131072 & Kroupa   &  8500 &  83438 & 1.0 & 0 & 10 & 64.57 & 100.9 & 30020 & $-1.02 \pm 0.01$  & $-0.30 \pm 0.03$ \\
9 & 131072 & Kroupa   &  8500 &  82777 & 1.0 &10 & 10 & 64.41 & 112.0 & 29525 & $-0.99 \pm 0.01$  & $-0.36 \pm 0.01$ \\[+0.1cm]
10 & 200000 & Kroupa   &  3500 & 127642 & 2.0 & 0 & 10 & 37.09 & 336.6 & 17950 & $-0.41 \pm 0.01$  & $0.43 \pm 0.03$  \\
11 & 200000 & Kroupa   &  5000 & 127523 & 4.0 & 0 & 10 & 52.14 & 937.9 & 31240 & $-0.78 \pm 0.01$  & $-0.47 \pm 0.02$ \\[+0.1cm]
12 & 131072 & Kroupa   &  5000 &  83281 & 2.0 & 0 &30 & 45.24 & 276.9 & 21750 & $-1.16 \pm 0.01$  & $-0.36 \pm 0.01$ \\
13 & 131072 & Kroupa   &  5000 &  83281 & 2.0 & 0 &50 & 45.24 & 276.9 & 22850 & $-1.16 \pm 0.01$  & $-0.36 \pm 0.01$ \\
14 & 131072 & Kroupa   &  5000 &  83853 & 2.0 & 0 &100 & 45.33 & 281.5 & 14450 & $-1.07 \pm 0.01$  & $-1.16 \pm 0.01$ \\[+0.1cm]
15 & 131072 & Kroupa   &  8500 &  83575 & 1.0 & 0 & 50 & 45.29 & 101.0 & 32400 &  $-1.13 \pm 0.01$ & $-0.45 \pm 0.03$ \\
16 & 131072 & Kroupa   &  8500 &  83438 & 1.0 & 0 &100 & 64.57 & 101.0 & 33150 & $-1.18 \pm 0.01$  & $-0.69 \pm 0.01$ \\
\end{tabular}
\end{table*}

The simulated clusters were moving on circular orbits through an external galaxy which was modeled as an isothermal sphere with constant
rotational velocity of $V_G=220$ km/sec. The distances of the clusters from the centers were chosen such that most clusters
had lifetimes of between 13 to 30 Gyr so that they are dynamically evolved to various degrees when they reach globular cluster ages. 
This resulted in initial Galactocentric distances of between 3.5 kpc to 8.5 kpc. 
Most clusters did not contain primordial binaries, however we also made one simulation of a star cluster starting with a primordial binary
fraction of 10\% to see how binaries influence our results. The binary stars in this cluster were set up as described
in \citet{lkb13}, with a flat period distribution between 1 and $10^6$ days and a thermal eccentricity distribution.

In addition to the $N$-body runs described above, we also used the best-fitting $N$-body models of individual globular clusters derived from the
simulations of \citet{baumgardt2017}. These models are based on $N$-body simulations of isolated star clusters starting with $N=100,000$ stars 
initially, which are scaled in mass and
radius to match the velocity dispersion and surface density profiles of Galactic globular clusters. The stars in these simulations 
followed a \citet{kroupa2001} IMF between mass limits $0.1 < m <100$ M$_\odot$ and a 10\% retention fraction was applied to all neutron stars
and black holes formed in the simulations. In this paper we only use the no-IMBH models of \citet{baumgardt2017}. 

Table~1 presents an overview of the performed $N$-body simulations. It gives for each simulation the number of cluster stars, the chosen 
initial mass function,
the Galactocentric distance of the cluster, the initial cluster mass, the primordial binary fraction, the assumed retention fraction of black holes,
the initial half-mass radius of the cluster, the initial tidal radius, the initial relaxation time and lifetime of the cluster 
(defined by the time a cluster has lost 99\% of its initial mass) and the best-fitting power-law slopes $\alpha$ of the global and 
central mass function at $T=12$ Gyr.

\section{Results}
\label{results_sec}

\subsection{Initial mass function}

We first compare the global cluster mass functions determined by \citet{sollimabaumgardt2017} with the evolution of 
clusters starting with different initial mass functions in order to determine the initial mass 
function of Galactic globular clusters.
For star clusters evolving in a tidal field, the shape of the global mass function changes as a result of
the preferential loss of low-mass stars
due to mass segregation and the removal of outer cluster stars \citep{vesperiniheggie1997, baumgardtmakino2003}.
If the cluster mass function is described by a power-law $N(m) \sim m^{\alpha}$, the
preferential loss of low-mass stars leads to an increase of the slope $\alpha$.\footnote{In the following we will
therefore speak of globular clusters with more negative values of $\alpha$ as being dynamically less evolved than clusters with larger values of $\alpha$.}

\begin{figure}
 \includegraphics[width=8.6cm]{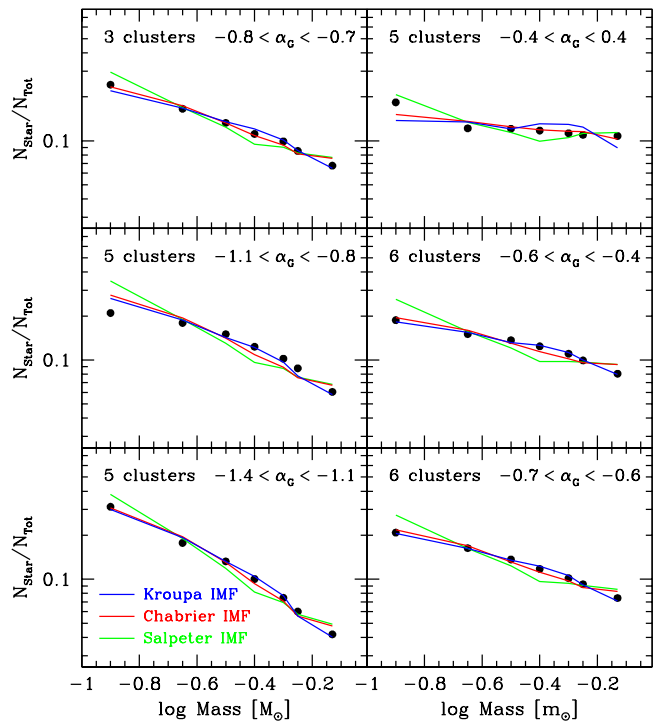}
\caption{Mass distribution of main sequence stars in globular clusters (circles) and simulations 4 to 6 from Table~1 (solid lines). Globular clusters are split into 6 different
groups depending on the power-law slope $\alpha_G$ of their global mass functions and the number of
stars is averaged over all clusters in each group. The number of globular clusters used is shown in the upper left corner of each panel. Solid lines show the corresponding
mass distributions of simulated star clusters starting with either a Kroupa (blue), Chabrier (red) or Salpeter mass function (green). 
Clusters starting with Kroupa or Chabrier mass functions provide a very good fit to the stellar 
mass distribution of globular clusters at each evolutionary stage, while the cluster starting with a Salpeter mass function provides a significantly worse
fit to the observed mass distribution.}
\label{noresidual}
\end{figure}

Fig.~\ref{noresidual} depicts the mass distribution of main sequence stars in globular cluster as determined by \citet{sollimabaumgardt2017}. 
Shown are the global mass functions which \citet{sollimabaumgardt2017} determined by fitting multimass King-Michie models to the observed mass functions.
We have split the cluster sample of \citet{sollimabaumgardt2017} into six different groups depending on the
best-fitting power-law slopes $\alpha_G$ of the global mass functions and show the average number of stars over all clusters in each group with solid circles.
The mass functions of globular clusters flatten due to the preferential loss of low-mass stars, from mass functions strongly increasing towards low-mass stars
in the lower left panel to flat mass functions in the upper right panel. In addition, the stellar mass function of the clusters in the three left panels
also show a flattening towards low-mass stars. Fitting a power-law only to the low-mass stars with $m<0.40$ in the lower left panel for example gives a
best-fitting slope $\alpha \approx -1$, while the higher-mass stars with $m>0.40$ are best fit by $\alpha \approx -1.4$.

The solid lines in Fig.~\ref{noresidual} depict the stellar mass distribution of clusters 4 to 6 from Table~1. The cluster starting with a Kroupa mass function is shown
by blue lines, while the clusters with Chabrier and Salpeter IMFs are shown by red and green lines respectively. All depicted clusters started with $N=131,072$
stars and had circular orbits at a Galactocentric distance of 5~kpc. For each cluster we calculate the global mass function slope $\alpha_G$ for each snapshot 
during the evolution and split the snapshots into the same six groups as the globular clusters. 
It can be seen that the stellar mass distribution in star clusters which start with either a Kroupa or Chabrier IMF is in very good agreement with the 
observed stellar mass distribution of globular clusters in each panel. In contrast, the cluster which starts with a (single power-law) 
Salpeter IMF overpredicts the number of low-mass stars with $\log m < -0.6$ and underpredicts the number of stars with  $\log m \approx -0.4$. These deviations are present for
the least evolved clusters in the lower left panel and are also present for the more evolved clusters. We obtain similar results for models 1 to 3
which start with $N=65536$ stars, as well as for clusters that start with a Kroupa IMF and $N=200,000$ stars and
take this as strong evidence that the mass function of globular clusters was not a single power-law at the low-mass end, but
had a steeper slope for high-mass stars than for lower-mass stars, with a break or turnover at around $m \approx 0.4$ M$_\odot$, in good
agreement to what is expected for either a Kroupa or Chabrier IMF.

\begin{figure*}
 \includegraphics[width=16cm]{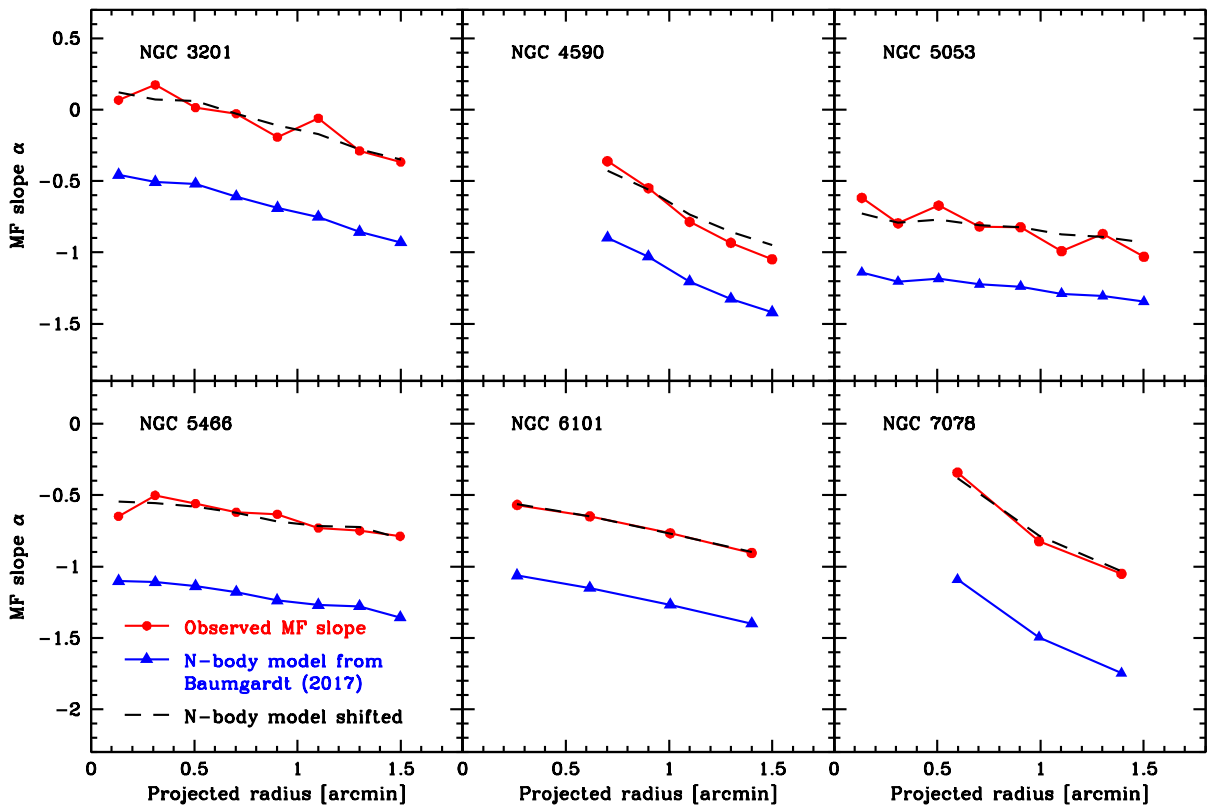}
 \caption{Mass function slopes in projection as a function of distance from the cluster centre for the six least evolved clusters from \citet{sollimabaumgardt2017}.
   Red, solid lines and circles show the observed MF slopes, blue solid lines and triangles show the best-fitting
    $N$-body models from Baumgardt (2017) for each cluster. Dashed lines show the MF slopes from the $N$-body simulations 
   shifted to correct for mass loss. The
   variation of the mass function slope with radius seen in the globular clusters can be entirely explained by two-body
     relaxation, indicating that the clusters started without primordial mass segregation.}
\label{sixcls}
\end{figure*}

In order to further explore the initial distribution of cluster stars, we depict in Fig.~\ref{sixcls} the observed mass function slope
as a function of distance from the cluster centre for the six
least evolved clusters from \citet{sollimabaumgardt2017}, (excluding NGC 6304 where the mass function is probably influenced by background contamination,
see discussion in Sollima \& Baumgardt 2017, and NGC 5024 where the mass function cannot be reliably determined in the centre due to incompleteness). 
We concentrate on the least evolved clusters since in the dynamically more evolved clusters the initial stellar distribution
has been strongly modified by the cluster evolution and ongoing dissolution and cannot be easily compared with the simulations
of \citet{baumgardt2017}. 
The observed mass functions are shown by red filled circles  and lines in Fig.~\ref{sixcls}. It can be seen that all depicted clusters are
mass segregated to various degrees since the mass function slopes decrease towards larger radii, implying a larger fraction of low-mass
stars at larger radii.
Also shown are the mass function slopes of the best-fitting $N$-body models from \citet{baumgardt2017} over the same
radial range for each cluster. The clusters in the simulations by \citet{baumgardt2017} are isolated clusters, hence the
global mass functions in these clusters do not evolve with time, unlike the mass functions in the observed clusters which change 
due to the preferential loss of low-mass stars. In order to account for this preferential loss of low-mass stars, we
shift the mass function slopes of each $N$-body model up until we obtain the best fit to the observed mass function slope for each
cluster. Except for this offset in the MF slope $\alpha$, 
the radial variation of the mass function is the same in the theoretical models and the observed clusters. Since the clusters in the
$N$-body simulations started unsegregated, the six depicted clusters must have also started without
primordial mass segregation and the radial variation of the mass function seen in each cluster is only due to energy equipartition driven
by two-body relaxation. Given the good agreement for the six depicted clusters, which span a large range of cluster masses and sizes, we conclude 
that most Galactic globular clusters started unsegregated. The only exception could be low-mass and low-density halo clusters like Pal~4,
which are highly mass segregated despite having long relaxation times and for which $N$-body simulations have shown that the amount of
mass segregation seen cannot be explained by dynamical evolution \citep{hasanizonoozietal2014,hasanizonoozietal2017}.

\subsection{Constraints on the cluster evolution}

\begin{figure}
 \includegraphics[width=8.6cm]{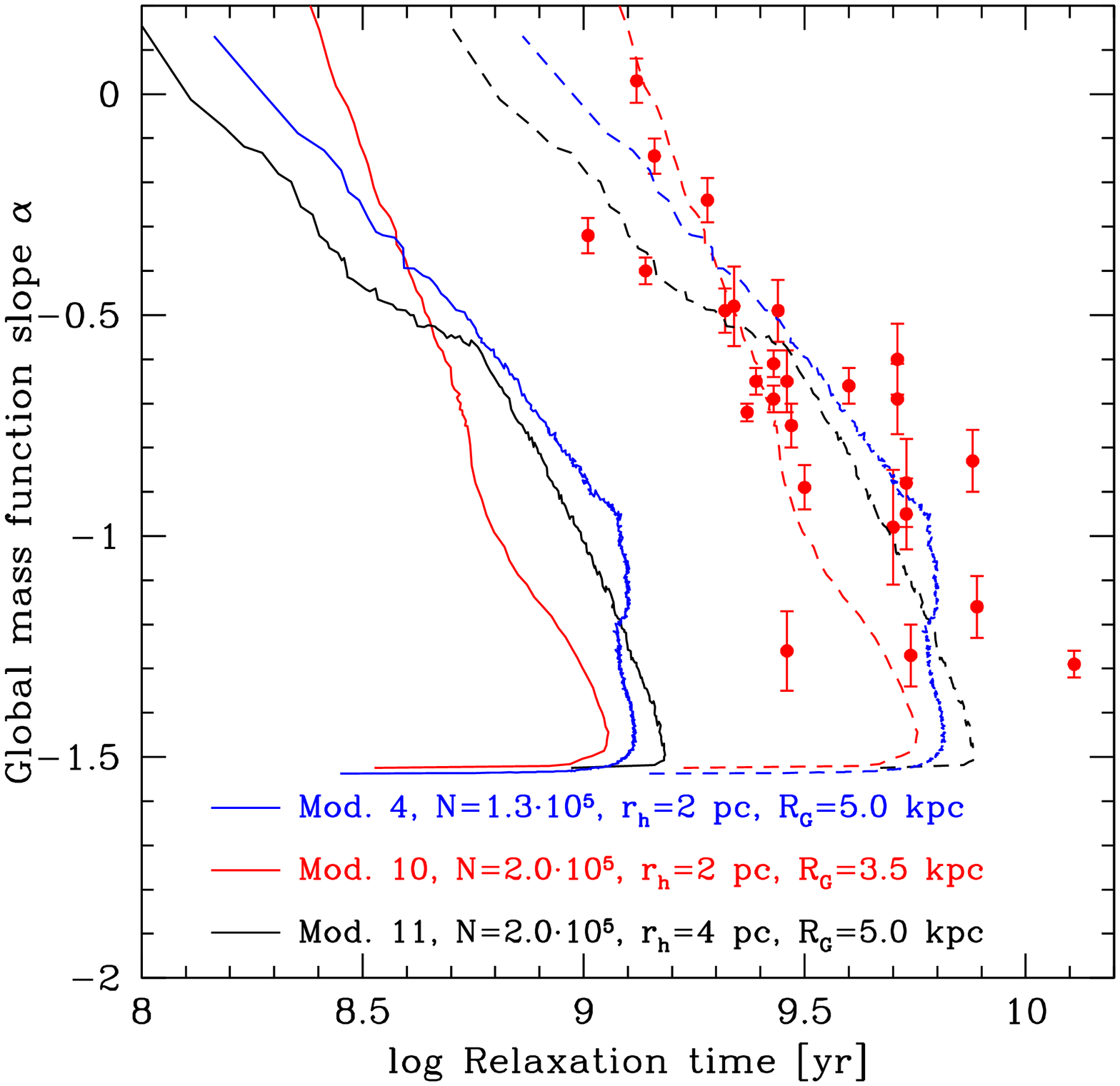}
 \caption{Global mass function slopes as a function of relaxation time for observed globular clusters (red circles) and three large $N$-body
  simulations of dissolving star clusters. Solid lines show the relaxation times of the simulated clusters, dashed lines show the
     simulated clusters after their relaxation times were multiplied by a factor 5 to account for the fact that galactic globular
     clusters are more massive and more extended than the clusters studied in the $N$-body simulations. After scaling there is a good overlap
      between both, indicating that Galactic globular clusters are tidally limited.}  
\label{trhalpha}
\end{figure}

\citet{sollimabaumgardt2017} found a tight anti-correlation between the half-mass relaxation time of globular clusters and their global mass
function slopes. 
Fig.~\ref{trhalpha} compares the location of the Galactic globular clusters in the relaxation time vs. mass function slope plane
which they found with results of three $N$-body simulations that start with a \citet{kroupa2001} IMF. Half-mass relaxation times $t_{rh}$ for
the clusters in the $N$-body simulations were calculated from eq. 2-62 of \citet{spitzer1987} using
all stars still bound to the clusters at any given time. We show the evolution of three $N$-body simulations starting with different
particle numbers, half-mass radii and Galactocentric distances but all having a low retention fraction of black holes.
The evolution of the clusters in the simulations can be divided into two phases, 
in the initial phase the clusters expand due to
stellar evolution mass loss and also heating due to binary stars in the core until they fill their Roche lobes.
In this phase the relaxation times increase but there is no strong evolution in the global mass function since the clusters
are not yet mass segregated, so clusters move to the right in the relaxation time vs. global MF slope plane. In the second phase, the
clusters have become mass segregated and their mass functions
become depleted in low-mass stars due to mass loss, so the MF slopes evolve towards more positive values. This two stage behavior
in the mass function evolution of star clusters was also found by \citet{lamersetal2013} and \citet{webbvesperini2016}. At the same time, the clusters
lose mass and shrink due to mass loss and a decreasing tidal radius, hence their relaxation times decrease as well. As a result, the simulated clusters
move from the lower right corner to the upper left in the relaxation time vs. global MF slope plane.

Since the clusters in the $N$-body simulations are about a factor 10 less massive and also on average about 30\% more compact than
observed globular clusters by the time they are 12 Gyr old, we increase the relaxation times of the star clusters in the $N$-body simulations
by a factor 5 to match the relaxation times of globular clusters and show these scaled curves by dashed lines in Fig.~3. 
The scaled $N$-body
clusters will probably not correctly capture the initial phases of cluster evolution, in particular the timescale for mass segregation and cluster 
expansion, but should describe the evolution of star clusters once they have become mass segregated and fill their Roche lobes since then
the evolution is driven mainly by a single process: mass loss.
It can be seen that the location of
the observed globular clusters agrees very well with that of the clusters in the $N$-body simulations when corrected for the differences in the
relaxation times, indicating that the anti-correlation between
mass function slope and relaxation time found by \citet{sollimabaumgardt2017} could be due to the ongoing dissolution of globular clusters. If correct,
Fig.~3 also implies that most globular clusters studied by \citet{sollimabaumgardt2017}
are tidally filling and their sizes start to shrink as they lose more and more of their stars. 
Such an evolution seems reasonable for many of the depicted clusters: A globular cluster with a mass of $M=3 \cdot 10^5$ M$_\odot$ 
orbiting at a distance of 4 kpc
from the Galactic centre for example would have a Jacobi radius of $r_J=60$ pc, and, assuming that $r_h/r_J \approx 0.10$ in the
tidally filling phase \citep{kuepperetal2008}, would have a half-mass radius of $r_h=6$ pc when tidally filling. Our simulated clusters
moving at $R_G=8.5$ kpc reach similar half-mass radii within a few Gyr, meaning that
many globular clusters, especially those in the inner parts of the Milky Way should also become tidally filling within a Hubble time. 
In addition, using the formula for globular cluster
lifetimes derived by \citet{baumgardtmakino2003}, we would expect the globular cluster described above to have undergone a 
significant amount of mass loss within 12 Gyr and have experienced significant changes to its initial mass function, similar to the
observed globular clusters.

\begin{figure}
 \includegraphics[width=8.6cm]{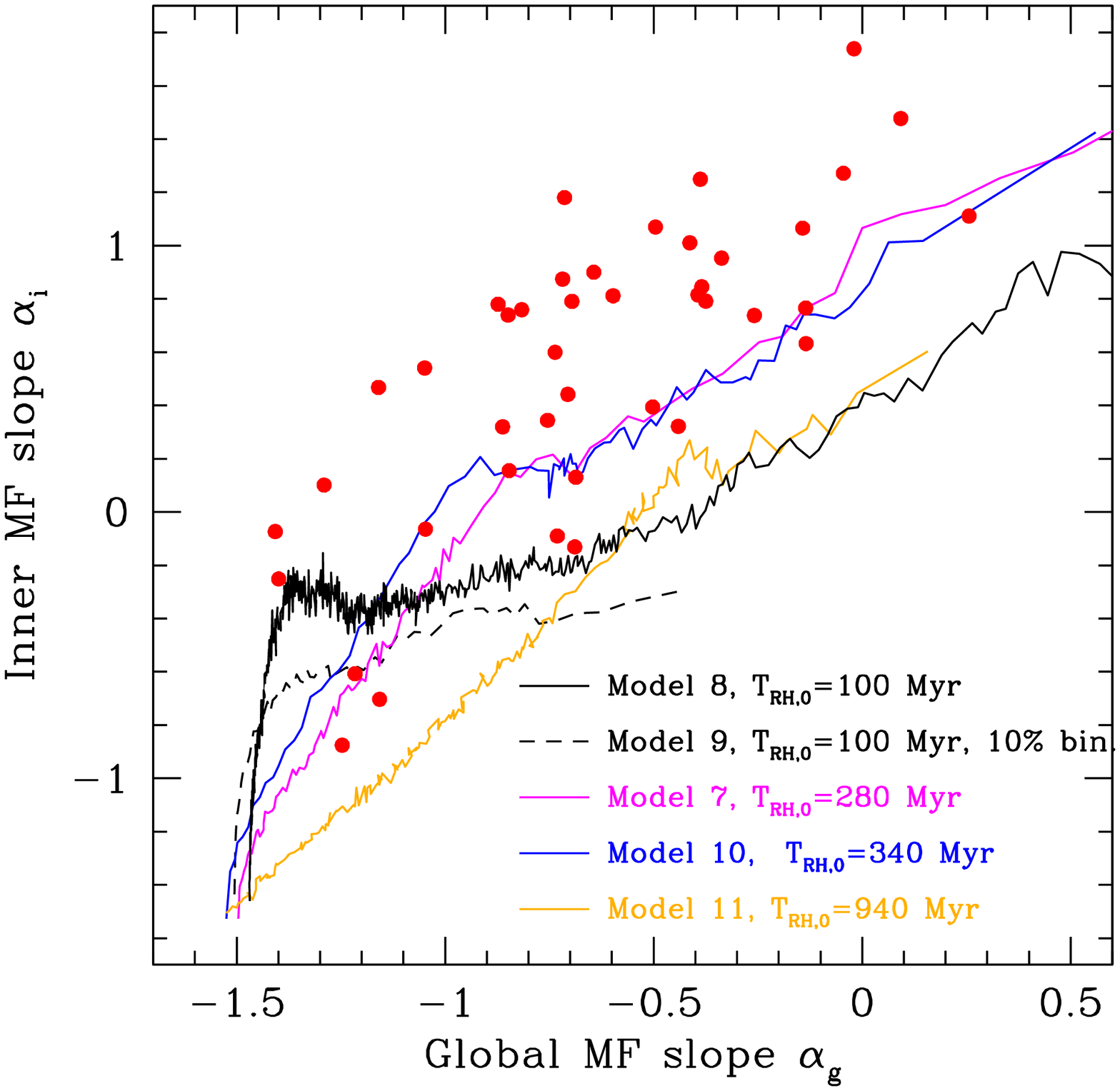}
 \caption{Global vs. inner mass function slope for observed globular clusters (red circles) and four different $N$-body simulations
   starting with a BH retention fraction of 10\% and different initial relaxation times.
The evolution of the simulated clusters falls into two categories, before core collapse
   when only the inner slope changes and the global slope stays nearly constant, and after core collapse when low-mass stars are preferentially lost
 and both slopes evolve. Observed globular clusters follow a very similar trend. The best match between observations and simulations is achieved
  for an initial relaxation time around 300 Myr.}
\label{figmfslope_bhlow}
\end{figure}

\begin{figure}
 \includegraphics[width=8.6cm]{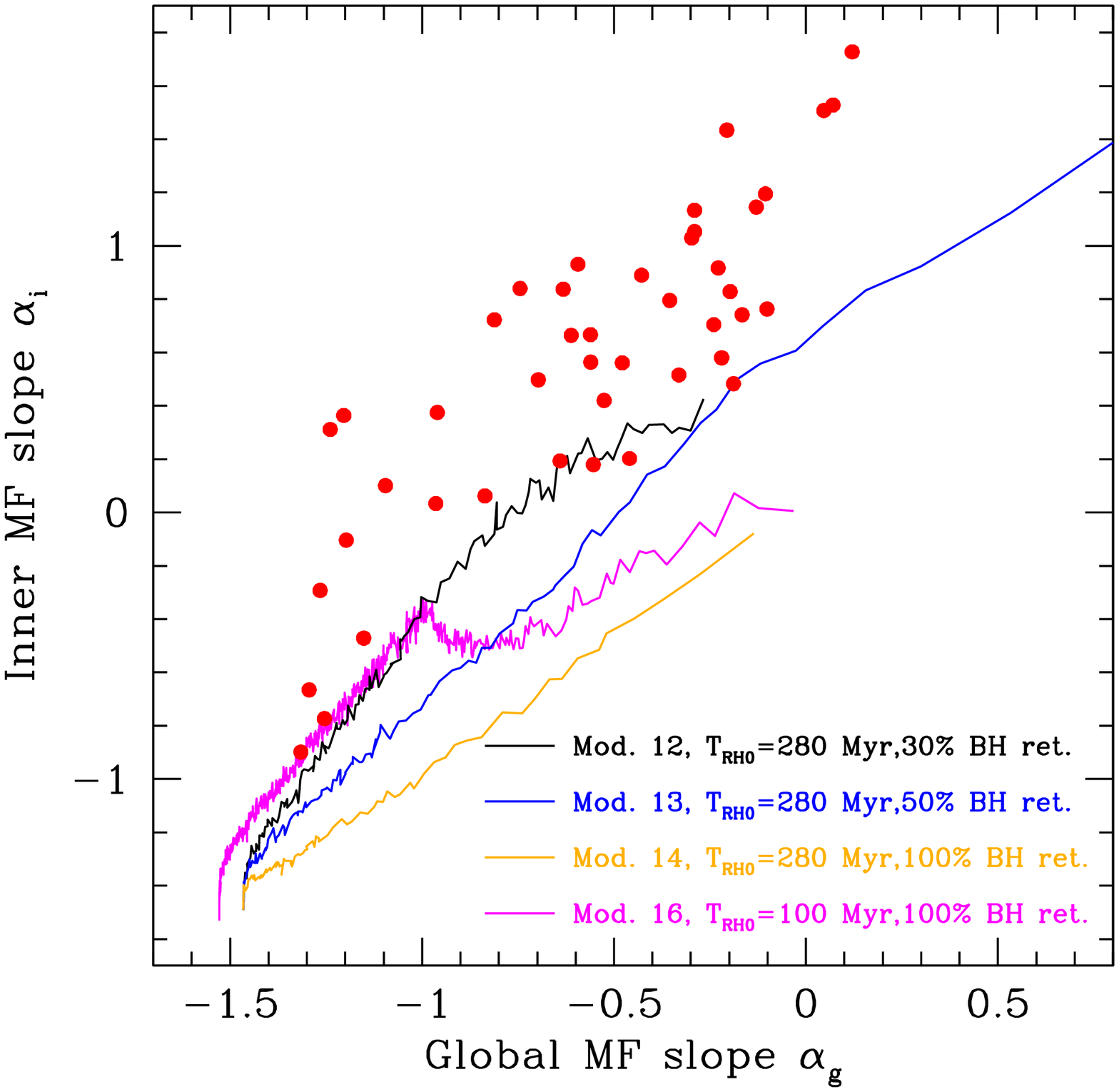}
 \caption{Same as Fig.\ \ref{figmfslope_bhlow} for clusters with a high initial BH retention fraction. It seems impossible to reproduce 
   the strong mass segregation seen in globular clusters with $N$-body models that have high stellar-mass BH retention fractions
    since the stellar-mass black holes suppress mass segregation among the low-mass stars.}
\label{figmfslope_bhhigh}
\end{figure}

Figs.~\ref{figmfslope_bhlow} and \ref{figmfslope_bhhigh} depict the location of the Galactic globular clusters in a global MF slope vs. inner MF slope plane.
Inner mass function slopes are derived from all stars located between projected radii of 0.15 $r_{hp} \le r \le$ 0.25 $r_{hp}$, 
where $r_{hp}$ is the projected half-light radius of a cluster. This radial range was chosen since in most clusters 
this is the innermost region where the completeness fraction is still higher than 80\% even for the faintest stars in the HST/ACS images.
For comparison, we also depict the evolution of the mass function slope for several $N$-body simulations. In Fig.~\ref{figmfslope_bhlow} we show
simulations with a 10\% retention fraction of stellar-mass black holes, while in Fig.~\ref{figmfslope_bhhigh} we depict clusters with
higher retention fractions.
The evolution of the clusters in the $N$-body simulations also falls into two phases. The clusters start with global and inner MF slopes
around $\alpha_g=\alpha_i \approx -1.5$ since they start from Kroupa IMFs without primordial mass segregation. Before core-collapse, clusters 
are not strongly mass segregated, therefore the global mass function changes only slowly, while the inner mass function evolves rapidly as the clusters become mass segregated. 
This near constancy of the global mass function before core collapse was also found by \citet{lamersetal2013} in $N$-body simulations.
After core collapse 
there is a strong evolution in both the inner and global mass function slope. Depending on the initial relaxation time of the clusters, 
core-collapse happens at slightly different points in the global vs. inner MF plane. Clusters with an initial relaxation time of $T_{RH,0} =100$ Myr
are already mass segregated before any mass loss has set in, while in clusters with $T_{RH,0} =900$ Myr core collapse takes
nearly as long as the dissolution of the clusters. Both curves seem incompatible with the location of the observed globular clusters. The best agreement 
with the location of observed globular cluster
is achieved for an initial relaxation time of $T_{RH,0} \approx 300$ Myr, implying an initial half-mass radius of $r_h=1$ pc for a $M_C=3 \cdot 10^5$ M$_\odot$ cluster.
Addition of 10\% primordial binaries leads only to a small change in the cluster evolution when all other parameters are kept the same
(see the evolution of model 8 vs. model 9). 

The mass function slopes of globular clusters also fall into two phases similar to the $N$-body clusters: a strong evolution in the 
inner MF slope together with a near constant
global MF slope until $\alpha_i \approx 0$, followed by a rapid evolution in both inner and global MF slope. In the latter
phase the difference between inner and global MF slope is nearly constant at $\alpha_i-\alpha_g \approx 1.1$
as clusters evolve towards dissolution. Given the
agreement between simulations and observations, we conclude that globular cluster mass functions are shaped by the same 
interplay of mass segregation and dissolution as the clusters in the $N$-body simulations. If correct, Fig.~\ref{figmfslope_bhlow} 
indicates that about 80\% of the clusters in the sample of \citet{sollimabaumgardt2017} 
have already undergone core collapse or are at least close to core collapse. In addition, globular clusters must have undergone a significant amount of mass-loss, since clusters 
in the $N$-body simulations which have lost half their initial cluster stars have a global MF slope of $\alpha_g=-1.0$.
Clusters in the simulations with $\alpha_g=-0.5$, which is a typical slope for the observed globular clusters, have already lost
75\% of their stars. Judging from Fig.~\ref{figmfslope_bhlow}, the most evolved globular clusters should already have lost 90\% 
of their initial cluster stars, and, for a constant mass loss rate,
should therefore dissolve within the next 1 to 2 Gyr. These estimates agree with the mass loss estimates obtained by \citet{wl2015},
who also concluded that globular clusters must have undergone a strong mass loss based on their present-day mass function slopes.  

\begin{figure}
 \includegraphics[width=8.6cm]{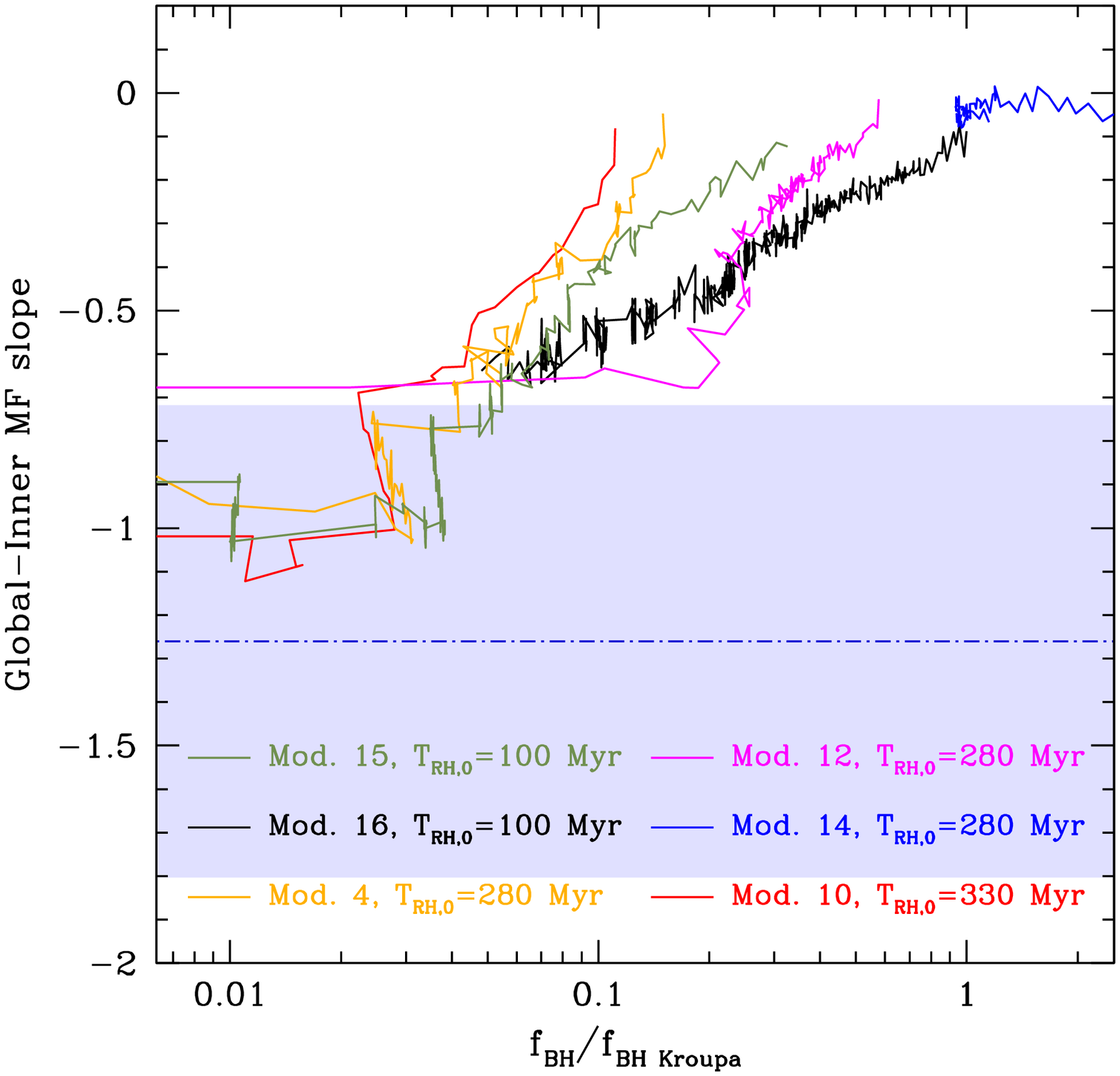}
 \caption{The difference between the global and the inner mass function slope for $N$-body simulations starting with different initial black hole retention fractions.
  All simulations shown start with a Kroupa mass functions and $N=131,072$ or $N=200,000$ stars. The mean difference for galactic globular cluster and its scatter are
   shown by the blue, dot-dashed line and the shaded area. Clusters in the $N$-body simulations reach a difference of $\alpha_g-\alpha_i \approx -1.1$
     only after nearly all black holes have been removed from the clusters.}
\label{figmfslopebh}
\end{figure}

Fig.~\ref{figmfslope_bhhigh} shows the evolution of star clusters with high BH retention fractions of 30\% to 100\%. Due to the high fraction
of stellar-mass black holes, mass segregation is strongly suppressed in these clusters and none of them shows strong mass segregation
for most of its evolution. This is also depicted in 
Fig.~\ref{figmfslopebh}, which compares the difference between the global and the inner mass function slope for our $N$-body simulations with the observed differences
for Galactic globular clusters. Galactic globular clusters have an average $\alpha_g-\alpha_i \approx -1.13 \pm 0.32$. The average $\alpha_g-\alpha_i$ and its 1$\sigma$
deviation are shown by the blue dashed-dotted line and the shaded area in Fig.~\ref{figmfslopebh}. The different curves in Fig.~\ref{figmfslopebh} show the evolution of
the difference of $\alpha_g-\alpha_i$ for six of the $N=131,072$ and $N=200,000$ star simulations from Table~1 which start with a Kroupa IMF. In most models
a decrease in the number of black holes is accompanied by an increase in the amount of mass segregation. Regardless of the assumed initial retention fraction of black holes,
the observed amount of mass segregation can only be reached when the fraction of black holes still retained in the clusters $N_{BH}/N_{BH Kroupa}$ is only a few percent 
of the fraction of all black holes that formed in the clusters. This poses a problem for the models which start with high initial retention fractions of either 50\% or
100\%. The only models which can reach large enough values $\alpha_g-\alpha_i$ are those that start with small initial relaxation times of $T_{RH}=100$ Myrs (e.g. model 16)
and even in this model the black holes are exhausted only very close to the end of the lifetime of the cluster. For larger initial relaxation times the clusters dissolve
before all black holes are ejected, hence mass segregation is either too slow (model 12) or no mass segregation is happening at all (model 14).
We conclude that the current number of black holes in globular clusters must be rather small, if clusters formed with a Kroupa IMF then at most
a few percent of the
initially formed black holes still remain in the clusters. For a typical globular cluster forming with $M=3 \cdot 10^5$ M$_\odot$, this
implies that no more than 50 stellar mass black holes currently reside in the cluster.

Given their masses, globular clusters probably started with relaxation times of several hundred Myr, more similar to our clusters starting with $r_h=2$ pc or $r_h=4$ pc initial
half-mass radius than the $r_h=1$ pc clusters. In addition, as we have seen before, most globular clusters have probably already lost a sizeable fraction of their
initial cluster mass, meaning that their lifetimes are of the order of 20 Gyr. A quick cluster dissolution together with the long initial relaxation times and
the small current black hole fraction is incompatible with a large initial black hole retention fraction. We therefore conclude that the initial black hole retention 
fraction in globular clusters was at most 50\%, otherwise it is impossible to explain the large amount of mass segregation seen in the clusters today.

\section{Discussion}

We have compared the observed stellar mass functions of 35 Galactic globular clusters recently determined by \citet{sollimabaumgardt2017}
from HST/ACS data with a set of large $N$-body simulations of star clusters dissolving in external tidal fields. We find that the observed mass functions are
compatible with globular clusters having started from either \citet{kroupa2001} or \citet{chabrier2003} mass functions but are incompatible with
\citet{salpeter1955} mass functions at the low mass end. Despite a difference
of up to $10^3$ in cluster mass, the IMF of globular clusters is therefore almost the same as that seen for stars in open clusters and field stars
in the Milky Way
\citep{bastianetal2010}. This is in agreement with theoretical star formation simulations which predict only a weak dependence of the shape of the
stellar mass function with environment \citep{myersetal2011, hennebelle2012}. The amount of mass segregation seen in the least evolved globular clusters 
can be completely explained by two body relaxation driven mass segregation. It therefore seems likely that 
globular clusters formed without primordial mass segregation at least among the low-mass stars with $m<0.8$ M$_\odot$.

The observations of \citet{sollimabaumgardt2017} have shown that the average
global mass function slope of globular clusters for stars with masses in the range $0.2 < m/M_\odot < 0.8$ is around $\alpha_g=-0.5$, higher by 1 
than the slope of the best-fitting power-law MF for a Kroupa mass function slope over the same mass range. According to our simulations, clusters that have global mass function slopes $\alpha_g=-0.5$ 
after a Hubble time have typical lifetimes of about 20 Gyr. Hence, for a constant mass loss rate, more than half of all globular 
clusters should dissolve within the next 10 Gyr.
From our simulations we also estimate that a typical globular cluster should have lost about
75\% of its initial stars and about 2/3 of its initial mass since formation. If globular clusters underwent an even more dramatic mass loss, as some 
scenarios used to explain the large fraction of 2nd generation stars in globular clusters imply \citep[e.g][]{antonacaloi2008,dercoleetal2008}, 
then this mass loss must have happened early on before globular clusters were significantly mass segregated. 

We also find a strong amount of mass segregation within globular clusters, the average difference between the global mass function slope to the
inner mass function slope (which we define as the mass function slope of stars around 20\% of the projected half-light radius) is about $\alpha_g - \alpha_i$ = -1.1.
Our simulations show that due to the effective suppression of mass segregation by stellar mass black holes, such a large amount of mass segregation
is only possible if the number of stellar mass black holes currently residing in the clusters is only a few percent of the initial number of black holes formed (for a Kroupa IMF).
A decrease in the amount of mass segregation or complete suppression of mass segregation due to stellar mass black holes has also 
been found previously by \citet{webbvesperini2016} and \citet{alessandrinietal2016}.
Our simulations show that clusters with black hole retention fractions equal to or higher than 50\% are not able to reach the low required black 
hole numbers before
final cluster dissolution unless their initial relaxation times would have been of order 100 Myrs or less. Such small relaxation times seem difficult to achieve for star
clusters starting with several $10^5$ M$_\odot$. We therefore conclude that the initial stellar mass black hole retention fractions were 50\%
or less.
This result is in agreement with \citet{sippeletal2013}, who found that the current number of BHs observed to
be in binary systems with a main-sequence companion as well as the estimated total number of BHs in M22 can be matched with a low initial
BH retention fraction of 10\%. Recently \citet{peutenetal2016} showed that the absence of mass segregation in NGC 6101 found by \citet{dalessandro2015}
can be explained by a high stellar-mass BH retention fraction. However, as Fig.~2 shows, the HST/ACS data actually shows NGC 6101 to be mass segregated, 
so there is currently no need to assume a high BH retention fraction in NGC 6101.

The clusters studied here only contain up to 200,000 stars initially and even though we do not find significant differences between the clusters with different
initial particle numbers studied here, it is not clear how our results scale to globular clusters which typically formed 
with a 5 to 10 times larger number of stars. For globular clusters in the inner parts of the Milky Way, where the tidal field
is strong, it seems possible that they could have expanded from small initial sizes to become tidally filling 
within a few Gyr and then undergo significant mass loss, e.g. undergo a similar evolution as the clusters in our simulations. Problems
could arise for clusters in the outer parts of the Milky Way, where the tidal field is too weak to allow expansion
up to the tidal radius and significant mass loss \citep{zonoozietal2011}. For such
clusters additional mass loss mechanisms, due to e.g. formation and evolution in a dwarf galaxy  \citep{webbetal2017}
or highly elliptic orbits \citep{hasanizonoozietal2017} might be necessary to create sufficient mass loss
to explain their present-day mass functions. Alternatively, we cannot completely rule out variations in the
global mass functions or primordial mass segregation in some globular clusters. Simulations of individual globular clusters on 
their exact orbits through the
Milky Way would help to further constrain their starting conditions, but are challenging since only low-mass or very extended globular clusters 
can be simulated with direct $N$-body simulations at the moment \citep{zonoozietal2011,heggie2014,wangetal2016},
while Monte Carlo codes can currently only handle constant tidal field strengths.

\section*{Acknowledgments}

We thank Elham Hasani Zonoozi and an anonymous referee for comments which helped improve the paper.

\bibliographystyle{mn2e}
\bibliography{mybib}

\label{lastpage}

\end{document}